# Lithographically defined site-selective growth of Fe filled multi-walled carbon nanotubes using a modified photoresist

Joydip Sengupta [a], Avijit Jana [b], N.D. Pradeep Singh [b], Chacko Jacob [a,*]

[a] Materials Science Centre, Indian Institute of Technology, Kharagpur 721302, India
[b] Department of Chemistry, Indian Institute of Technology, Kharagpur 721302, India

ABSTRACT

Partially Fe filled multi-walled carbon nanotubes (MWCNTs) were grown by chemical vapor deposition with propane at 850 oC using a simple mixture of iron (III) acetylacetonate (Fe(acac)$_3$) powder and conventional photoresist. Scanning electron microscopy revealed that catalytic nanoparticles with an average diameter of 70 nm are formed on the Si substrate which governs the diameter of the MWCNTs. Transmission electron microscopy shows that the nanotubes have a multi-walled structure with partial Fe filling. A site-selective growth of partially Fe filled MWCNTs is achieved by a simple photolithographic route.

Carbon nanotubes (CNTs) exhibit a unique combination of electronic, thermal, mechanical and chemical properties, which promise a wide range of potential applications in key industrial sectors. In the last few years, chemical vapor deposition (CVD) has been the preferred method among different CNT growth methods because of its potential advantage to produce a large amount of CNTs growing directly on a desired substrate with high purity, large yield and controlled alignment. The CVD synthesis of CNTs on plain substrates generally requires catalyst metal deposition over the substrate, which is time consuming and the deposited area is also finite. These limitations can be easily overcome by employing the spin-coating of the catalyst material on the desired substrate. Moreover, in recent times, magnetic metal encapsulated CNTs are of huge interest as it has demonstrated widespread applications in magnetic force microscopy [1], high-density magnetic recording media [2] and biology [3]. However, there are many technical barriers to achieve magnetic metal encapsulated CNT-based electronic devices; especially the placement of CNTs in the desired position is crucial for CNT integration in such devices. In this regard, photolithography is one of the most important and powerful patterning techniques for microfabrication and is also widely used in integrated circuit (IC) technology. Thus we have employed spin-coating along with the lithographic approach for the site-selective growth of magnetic material filled CNTs using a simple mixture of iron (III) acetylacetonate (Fe(acac)$_3$) powder and conventional photoresist (HPR 504, Fuji Film) as a starting material.

* Corresponding author: Fax: +91 3222255303.
E-mail address: cxj14_holiday@yahoo.com (C. Jacob).

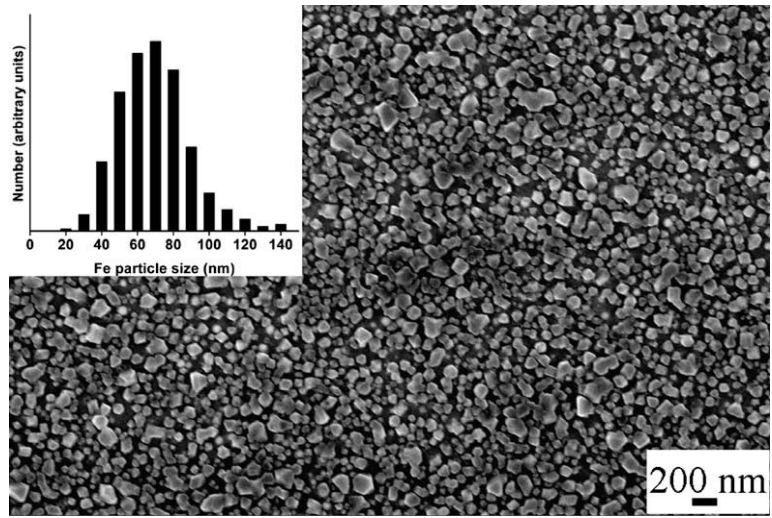

Fig. 1 – SEM micrograph of the catalytic nanoparticles prepared from Mod-PR film over the Si(1 1 1) substrate after annealing at 900 oC; (inset) the distribution histogram of the catalyst nanoparticles.

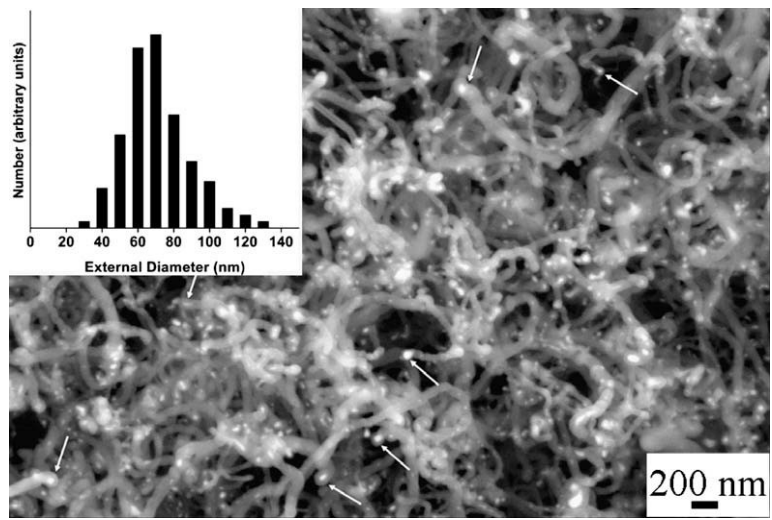

Fig. 2 – SEM micrograph of the MWCNTs synthesized using Mod-PR by CVD growth; (inset) the distribution histogram of the external CNT diameter deposited by the CVD method.

706 mg of Fe(acac)$_3$ was mixed with 10 ml of HPR 504 to obtain a modified photoresist (Mod-PR) solution of concentration 0.2 M. Then the Mod-PR solution of concentration 0.2 M was spin-coated with a rotation speed of 4000 rpm for 20 s on the Si(1 1 1) substrate to get a thin layer of the Mod-PR. The thin Mod-PR film was annealed in air for 10 min at 200 oC to improve the adhesion to the substrate. The substrates were then loaded into a quartz tube furnace (ELECTROHEAT EN345T), pumped down to $10^{-2}$ Torr and backfilled with flowing argon to atmospheric pressure. Afterwards, the samples were heated in argon up to 900 oC following which the argon was replaced with hydrogen. Subsequently, the samples were annealed in hydrogen atmosphere for 10 min. Finally, the reactor temperature was brought down to 850 oC and the hydrogen was turned off, thereafter propane was introduced into the gas stream at a flow rate of 200 sccm for 1 h for CNT synthesis.

For lithographically selective growth of CNTs, after performing the spin-coating of Mod-PR the samples were baked at 90 oC for 15 min followed by an exposure step with a mask aligner to make an array of patterns. The exposed specimens were developed in the developer solution for 60 s and rinsed in distilled water. Thereafter, the synthesis of CNT was carried out on the pattern areas following the same procedure as described earlier.

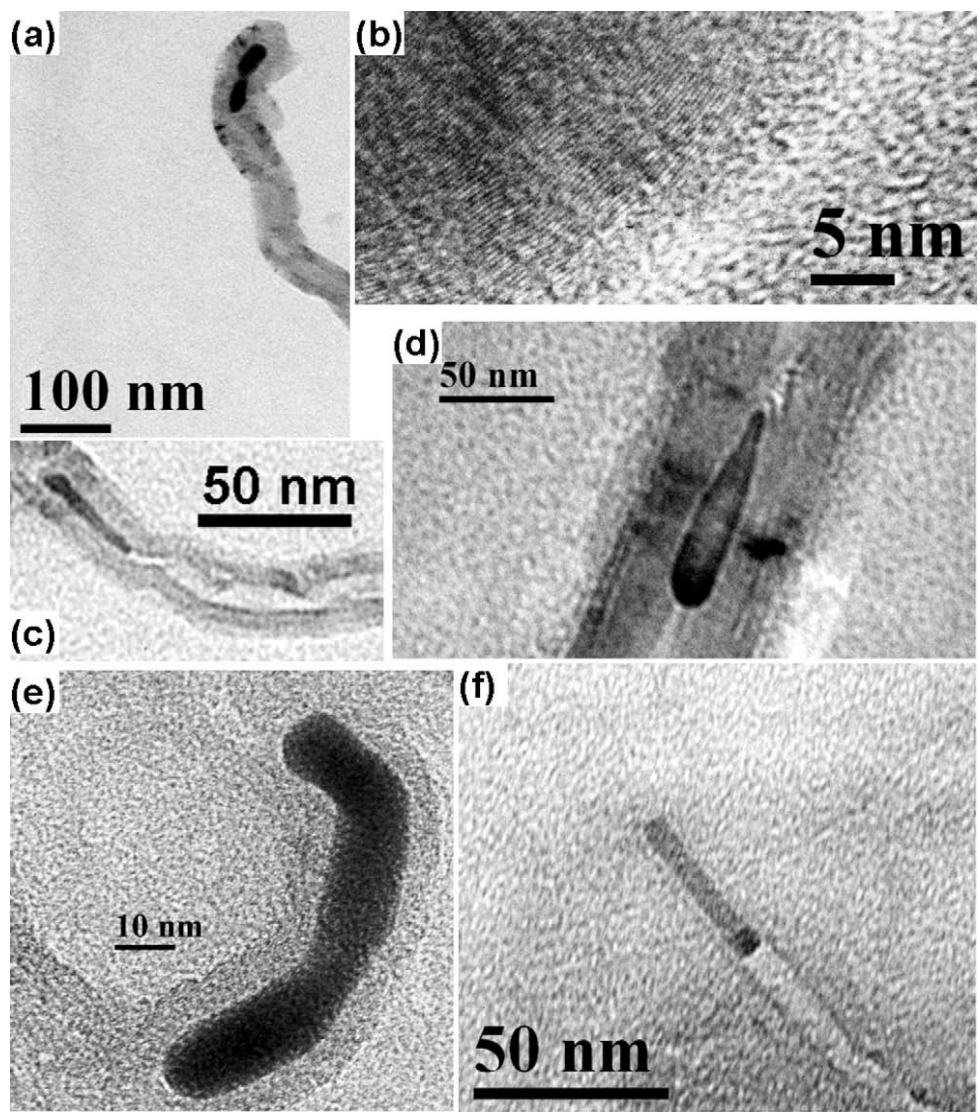

Fig. 3 – TEM images of the Fe encapsulated CNTs grown by the CVD method using Mod-PR, (a) A catalytic nanoparticle is encapsulated at the top of a CNT implying tip-growth mechanism; (b) TEM image of interlayer spacing of polyaromatic carbon in CNT; (c–f) Fe nanoparticles encapsulated in CNTs.

Fig. 1 shows the scanning electron microscopy (SEM) image of the catalytic nanoparticles formed over the Si(1 1 1) substrate after annealing the Mod-PR film at 900 oC. The statistical analysis of the size of the catalytic nanoparticles (inset of Fig. 1) displays a Gaussian profile with a mean of 70 nm. Fig. 2 shows the SEM image of the CNTs synthesized on the Si(1 1 1) substrate by CVD using Mod-PR. The high-aspect-ratio nanostructures had a randomly oriented spaghetti-like morphology with high area density. We note that in Fig. 2 the outer diameters of CNTs are in the range of 30–130 nm and the corresponding distribution profile (inset of Fig. 2) is similar to the size distribution profile of the catalyst nanoparticles (inset of Fig. 1). Thus, the size of the catalytic particles largely determines the outer diameters of the CNTs. Several bright catalyst particles were detected at the tip of the tubes (indicated in Fig. 2 with white arrows) suggesting the tip-growth mechanism.

Fig. 3 shows the transmission electron microscopy (TEM) images of MWCNTs grown using Mod-PR. A catalytic nanoparticle of size nearly 80 nm is encapsulated at the tip of a CNT (Fig. 3a) implying tip-growth mechanism. It can be clearly seen that the CNT is well graphitized and have clear well-ordered lattice fringes of polyaromatic sheets (Fig. 3b). Fig. 3(c–f) demonstrate the presence of high-aspect-ratio nanoparticles encapsulated inside the CNTs deposited using Mod-PR. Chemical composition analysis confirms that the elongated particles are of Fe (not shown here). The growth model of partially Fe filled CNT can be explained by emphasizing the role of the capillary action of the liquid-like Fe particles that exist at the time of CNT nucleation. The detail of the growth model is discussed elsewhere [4].

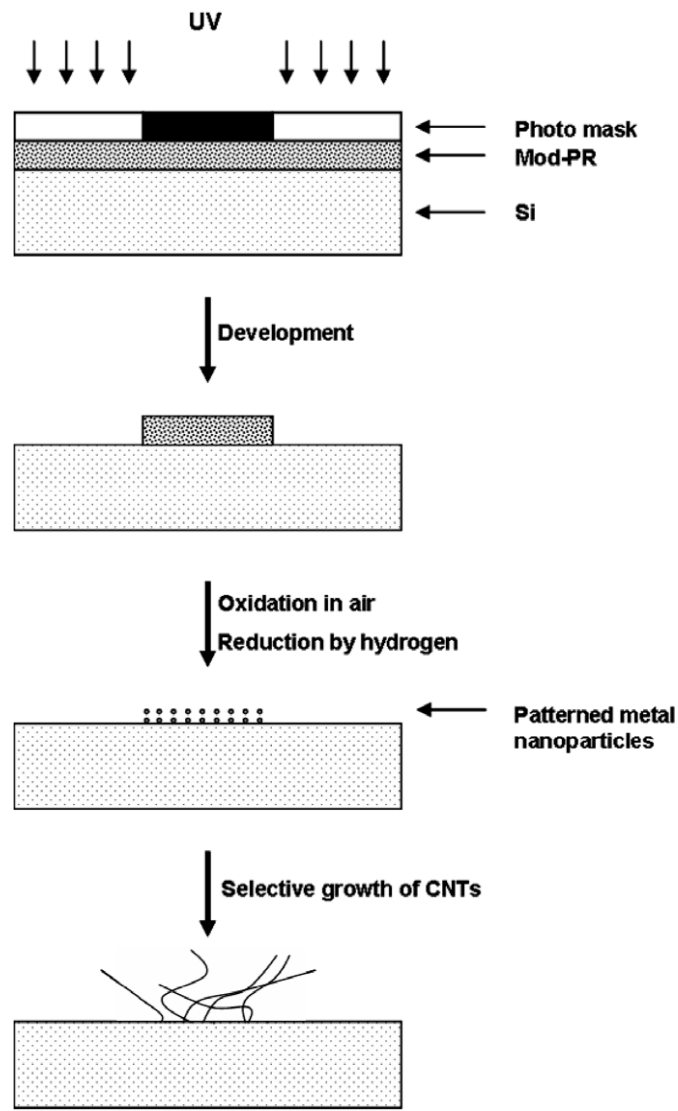

Fig. 4 – Schematic representation of the direct photolithographic route for the selective growth of partially Fe filled MWCNTs.

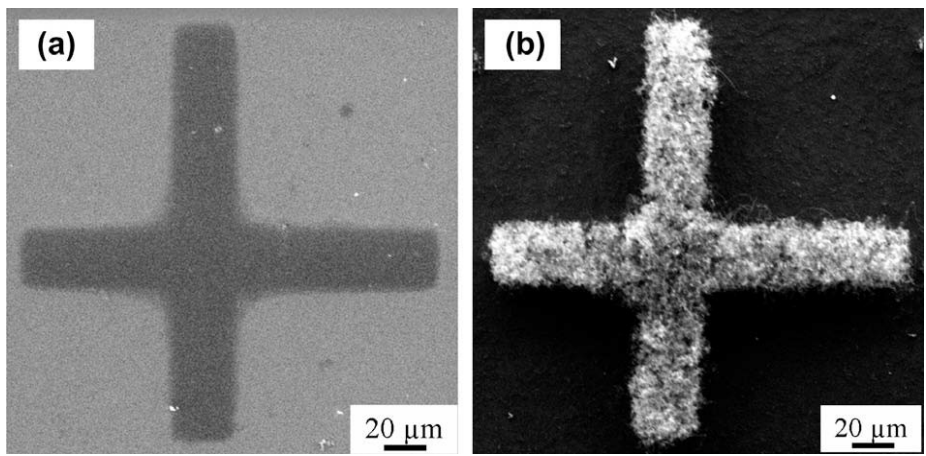

Fig. 5 – Site-selective growth of partially Fe filled MWCNTs by direct photolithographic route using Mod-PR (a) SEM image of Mod-PR pattern; (b) SEM image after CNT growth.

Fig. 4 shows the schematic representation of the direct photolithographic route adopted for the selective growth of partially Fe filled MWCNTs. Fig. 5a is the SEM image of the Fe-containing patterns produced by the general photolithographic process using Mod-PR whereas, Fig. 5b is the SEM image of the CNTs grown from lithographically defined catalyst patterns. This photolithography-based approach is simple and yet robust process with excellent potential to fabricate partially Fe filled CNTs at lithographically defined locations in a reproducible manner.

In conclusion, we have demonstrated a simple and effective method to synthesize partially Fe filled CNTs using a mixture of Fe(acac)$_3$ powder and conventional photoresist and successive CVD growth. The experimental results show that the catalyst particle size effectively controls the diameter of the partially Fe filled CNTs grown through a tip-growth mechanism. A simple photolithographic route has been devised for the spatially selective synthesis of partially Fe filled MWCNTs, which were grown only on a pre-defined surface in a manner compatible with current IC technology.

# Acknowledgement

J. Sengupta is thankful to CSIR for providing the senior research fellowship.

# REFERENCES

[1] Winkler A, Mühl T, Menzel S, Koseva RK, Hampel S, Leonhardt A, et al. Magnetic force microscopy sensors using iron-filled carbon nanotubes. J Appl Phys 2006;99:1049051–5.
[2] Kuo CT, Lin CH, Lo AY. Feasibility studies of magnetic particle-embedded carbon nanotubes for perpendicular recording media. Diamond Relat Mater 2003;12:799–805.
[3] Palen EB. Iron filled carbon nanotubes for bio-applications. Mater Sci (Poland) 2008;26:413–8.
[4] Sengupta J, Jacob C. The effect of Fe and Ni catalysts on the growth of multiwalled carbon nanotubes using chemical vapor deposition. J Nanopart Res. 2010;12:457–65.